\documentclass[a4paper, amsfonts, amssymb, amsmath, reprint, showkeys, nofootinbib, twoside, superscriptaddress, floatfix]{revtex4-1}
\usepackage[english]{babel}
\usepackage[utf8]{inputenc}
\usepackage[colorinlistoftodos, color=green!40, prependcaption]{todonotes}
\usepackage{amsthm}
\usepackage{mathtools}
\usepackage{physics}
\usepackage{xcolor}
\usepackage{graphicx}
\usepackage[left=23mm,right=13mm,top=35mm,columnsep=15pt]{geometry} 
\usepackage{adjustbox}
\usepackage{placeins}
\usepackage[T1]{fontenc}
\usepackage{lipsum}
\usepackage{csquotes}
\usepackage[pdftex, pdftitle={Article}, pdfauthor={Adrian L. Hauber}]{hyperref} % For hyperlinks in the PDF
\usepackage{cleveref}
\usepackage{color,soul}

\newcommand{\nsurr}{500}
\newcommand{\nreal}{n}

\begin{document}

\title{Detecting frequency modulation in stochastic time series data}

\author{Adrian L. Hauber}
    \email[Correspondence email address: ]{adrian.hauber@fdm.uni-freiburg.de}
    \affiliation{Institute of Physics, University of Freiburg, 79104 Freiburg, Germany}
    \affiliation{Freiburg Center for Data Analysis and Modeling (FDM), University of Freiburg, 79104 Freiburg, Germany}

\author{Christian Sigloch}
    \affiliation{Developmental Biology, Institute Biology I, Faculty of Biology, University of Freiburg, 79104 Freiburg, Germany}
    \affiliation{Spemann Graduate School of Biology and Medicine (SGBM), University of Freiburg, 79104 Freiburg, Germany}

\author{Jens Timmer}
    \affiliation{Institute of Physics, University of Freiburg, 79104 Freiburg, Germany}
    \affiliation{Freiburg Center for Data Analysis and Modeling (FDM), University of Freiburg, 79104 Freiburg, Germany}
    \affiliation{Centre for Integrative Biological Signalling Studies (CIBSS), University of Freiburg, 79104 Freiburg, Germany}

\date{\today} % Leave empty to omit a date

\begin{abstract}
We propose a new statistical test to identify non-stationary frequency-modulated stochastic processes from time series data. Our method uses the instantaneous phase as a discriminatory statistics with reliable critical values derived from surrogate data. We simulated an oscillatory second-order autoregressive process to evaluate the size and power of the test. We found that the test we propose is able to correctly identify more than 99$\%$ of non-stationary data when the frequency of simulated data is doubled after the first half of the time series. Our method is easily interpretable, computationally cheap and does not require choosing hyperparameters that are dependent on the data.  
\end{abstract}


\maketitle

\section{Introduction} \label{sec:introduction}
Methods to acquire biology data on the cellular level are becoming increasingly relevant \cite{codeluppi_spatial_2018,moffitt_molecular_2018,svensson_exponential_2018,stuart_integrative_2019}. Their interpretation, however, may be difficult because data captured at the level of a single cell frequently are of stochastic nature. This is especially important for time-resolved data where repeating the measurement is often not feasible either from a technical or biological point of view.  

%"Narrower problem"
Oscillations in the concentration of cellular components such as proteins have been shown to be of great importance, e.g.\ in the development of organs such as the brain \cite{sueda_high_2019}. Identification and analysis of such oscillations is therefore a crucial part of understanding regulatory biological systems that govern development. %\cite{sigloch_2021}. 

%"Yet narrower paper gap"
Especially the frequency of oscillations is an important regulator for many processes, e.g.\ the development of vertebrae that is tuned via Delta-Notch inter-cellular signalling \cite{liao_faster_2016,harima_accelerating_2013}. Given oscillatory data, the question arises whether oscillations have a constant or modulated frequency, i.e.\ whether the underlying stochastic process is stationary or non-stationary. Here, we propose a new method that is specifically tailored to detect frequency modulations in short time series in contrast to existing approaches for identifying non-stationarity \cite{dickey_distribution_1979,kwiatkowski_testing_1992, zivot_further_2002,phillips_testing_1988,elliott_efficient_1996,timmer_power_1998-2,mohanty_robust_2000}.

%>> "Summary"
%"Approach"
We employ the Hilbert transform to obtain instantaneous phase information from a time series to track the speed with which oscillations proceed. We compare this to the instantaneous phase of surrogate time series with constant frequency to derive confidence bands and to identify frequency modulations.

%"Results"
To investigate the size and power of this test, we simulated an oscillating second-order autoregressive process with different degrees of frequency modulation. We applied our proposed method and counted how often the test identifies a time series as non-stationary depending on the actual frequency modulation.

We found that the test we propose is able to correctly identify more than $99\%$ ($97\%$) of the frequency modulated time series when the frequency is increased by $100\%$ ($80\%$) with respect to the base frequency.

%We also applied our method in a biological context in order to identify frequency-modulated time series \cite{sigloch_2021}.% and found that it rejected the null hypothesis of stationarity for a seemingly non-stationary time series.

We also applied our method to a time series from a biological experiment in order to identify frequency-modulated time series.

\section{The test} \label{sec:develop}

\subsection{Instantaneous phase as a measure for stationarity}
\label{sec:instphase}

%Outline
%\begin{itemize}
%\item time series can be non-stationary, e.g.\ by exhibiting a time-dependent frequency (\Cref{fig:dataphase}a)
%\item instantaneous phase increases as the oscillation proceeds (\Cref{fig:dataphase}b)
%\item instantaneous phase shows a kink when frequency switches from one value to another
%\end{itemize}

Non-stationarity can for example be exhibited by a time-dependent frequency of oscillation. % (\cite{sigloch_2021}). 
The instantaneous phase allows to trace oscillations and thereby detect changes in the frequency, i.e.\ the speed with which oscillations proceed.

%Non-stationarity
%\begin{itemize}
%\item time series can be thought of as a realization of a stochastic process
%\item process is called stationary if its distribution function is time invariant
%\item non-stationarity can come in different manifestations, e.g.\ time-dependent frequency which we focus on in this manuscript (\Cref{fig:dataphase}a)
%\item because in neural development, there is evidence that the oscillatory expression of the Her6 protein changes its frequency from one value to another when proceeding to a different developmental stage (???)
%\item time series that are realizations of non-stationary processes will be called non-stationary time series in this manuscript
%\end{itemize}

Time series can be thought of as realizations of stochastic processes. Stationary stochastic processes have distribution functions that are shift-invariant in time. The second-order autoregressive (AR[2]) process with variance $\sigma^2$,
\begin{align}
    x(t) = a_1 x(t-1)+a_2 x(t-2) + \epsilon(t), \, \epsilon \sim \mathcal{N}(0,\sigma^2),
    \label{eq:ar2process}
\end{align}
is an example for a stationary linear stochastic process that can oscillate. The parameters $a_1,a_2$ can be calculated from the period $T=1/\nu$ and the mixing time $\tau$ \cite{timmer_cross-spectral_1998}, which is the decay time of the auto-correlation function, by 
\begin{align}
    a_1&=2 \cos(2\pi \nu)\,\mathrm{e}^{-1/\tau}
\label{eq:a1}
 \\ 
a_2 &= -\mathrm{e}^{-2/\tau}.
\label{eq:a2}
\end{align}

\begin{figure}
    \includegraphics[width = 8.6cm]{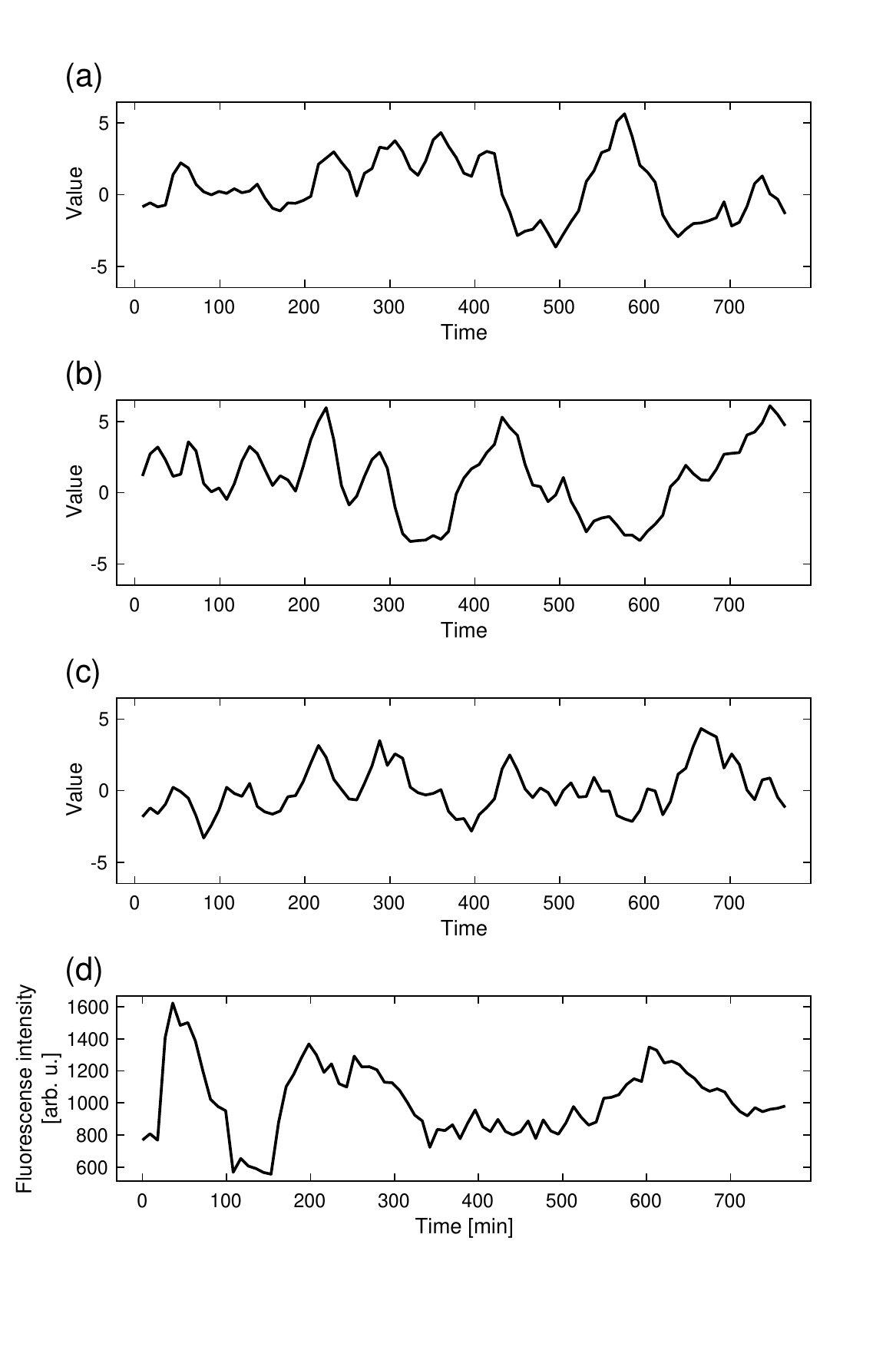}
    \caption{Three realizations of an oscillating second-order autoregressive process (\Cref{eq:ar2process}) with $\sigma^2 = 1$ and (a) $T=270$, (b) $T=180$, (c) $T=90$, as well as $\tau=22.5$ and a sampling interval of $9$ corresponding to the experimental data in (d). (d) Experimental time series of HES/Her family protein expression in neural stem cells of zebrafish larvae, captured using light sheet microscopy.}% (data taken from \cite{sigloch_2021}, \textit{e15c064}).}
    \label{fig:simstudy_arpars}
\end{figure}

Given simulations of this stationary stochastic process (\Cref{fig:simstudy_arpars}a-c), it is hard to decide by visual inspection whether an experimental time series (\Cref{fig:simstudy_arpars}d) is a realization of a stationary process. 

Non-stationarity can appear in different manifestations, e.g.\ a time-dependent frequency which we focus on here. We observed this property in the time series of fluorescence intensities of HES/Her family protein expression in neural stem cells located in the thalamic proliferation zone of zebrafish larvae. It is important to note however that there are many more properties that can change over time due to the underlying stochastic process being non-stationary. For the sake of simplicity, we will skip the reference to the underlying stochastic process and denote the time series itself as ``stationary'' or ``non-stationary''.

%Instantaneous phase
%\begin{itemize}
%\item frequency $\nu$ is usually defined by the number of oscillations per unit time
%\item van der Pol defines frequency as the derivative of a quantity called instantaneous phase which allows to extend the concept of frequency to non-stationary signals \cite{van_der_pol_fundamental_1946}
%\item for a harmonic oscillator, $x(t) = A\cos(2\pi \nu t + \psi)$, instantaneous phase is the argument of the cosine and is a linear function of time
%\item a stochastic process that depicts stationary oscillations has a approximately linearly increasing phase (\Cref{fig:dataphase}b for $0<t<300$) 
%\item when frequency is a function of time, phase will deviate from linear dependence on time, e.g.\ have kink for two step-function frequency (\Cref{fig:dataphase}b at $t=300$)
%\item therefore, instantaneous phase can be used to check if frequency is time-dependent
%\end{itemize}

Frequency $\nu$ is usually defined as the number of oscillations per unit time. Van der Pol however defined frequency as the derivative of a quantity called instantaneous phase $\Phi(t)$ \cite{van_der_pol_fundamental_1946}
\begin{align}
\nu = \frac{1}{2\pi} \frac{\mathrm{d}\Phi(t)}{\mathrm{d}t},
\end{align}
which allows to extend the concept of frequency to time series that have a time-dependent frequency or are ``frequency-modulated''. For example, a harmonic oscillator $x(t) = A\cos(2\pi \nu t + \psi)$ with $A,\nu,\psi = \mathrm{const.}$ has an instantaneous phase of $\Phi(t) = 2\pi \nu t + \psi$, which is a linear function of time. 

\begin{figure}
\includegraphics[width = 8.6cm]{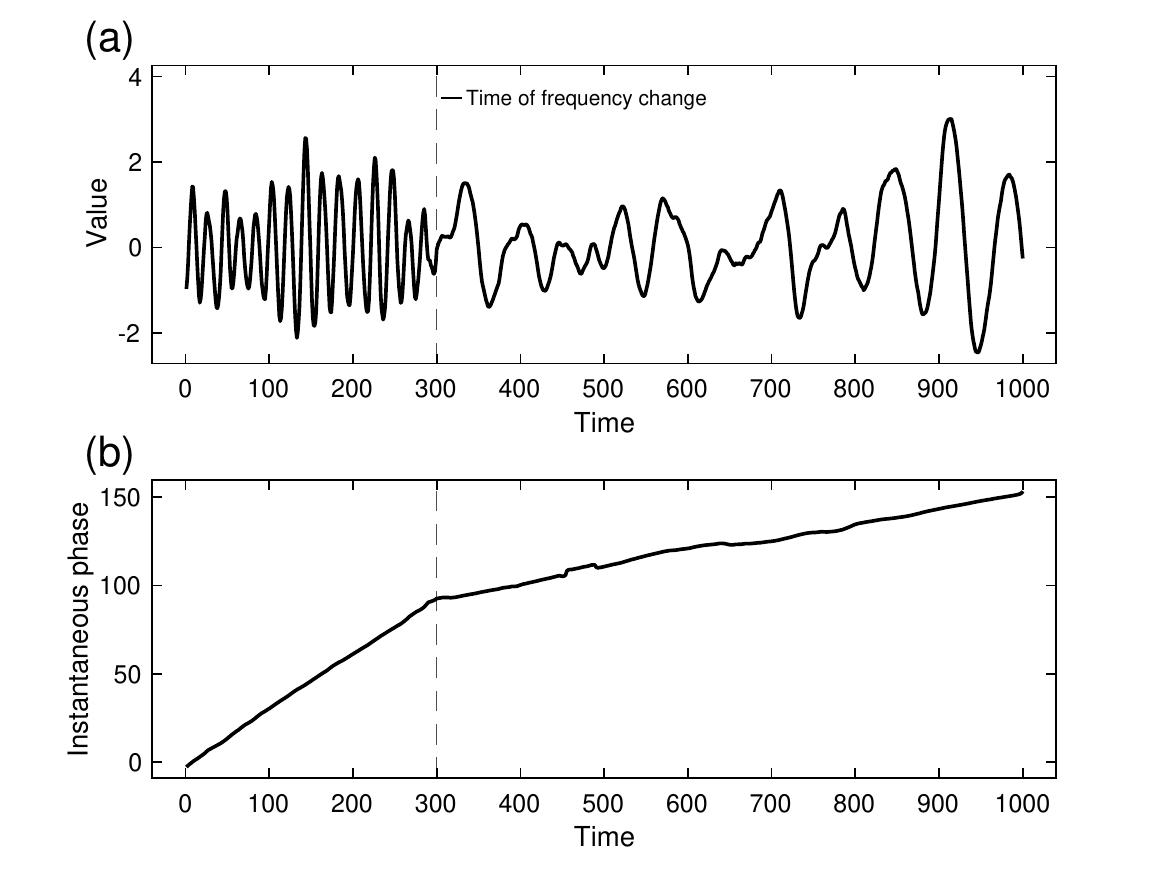}
\caption[Example data and instantaneous phase]{(a) Example non-stationary time series with a constant high frequency for $t<300$ and a constant low frequency for $t\geq300$. (b) Unwrapped instantaneous phase of the time series in (a). The kink in the instantaneous phase is a consequence of the non-stationarity.}
\label{fig:dataphase}
\end{figure}

For a stochastic process that depicts stationary oscillations, the instantaneous phase is also approximately linear in time (\Cref{fig:dataphase}b for $0<t<300$). When frequency is modulated e.g.\ with a step-function, the instantaneous phase will deviate from the linear appearance and have a kink at the point in time where the adjustment takes place (\Cref{fig:dataphase}b at $t=300$). 

%Hilbert-transform
%\begin{itemize}
%\item the analytical signal $x^{(a)}(t) = x(t) + \mathrm{i}\tilde{x}(t)$ provides a complex representation of a real-valued time series \cite{gabor_theory_1946}
%\item the instantaneous phase can be calculated as the complex argument of the analytical signal
%\item the imaginary part of analytical signal is calculated by imposing a phase shift of $\pi/2$ to the real-valued time series, which is achieved by applying a discrete Hilbert-transform
%\item which can be obtained from the Fourier transform by neglecting negative frequency components and back-transforming into the time-domain \cite{marple_computing_1999}
%\end{itemize}

For a given time series $x(t)$, the instantaneous phase can be calculated as the complex argument of the analytic signal. The analytic signal $x^{(a)}(t) = x(t) + \mathrm{i}\tilde{x}(t)$ provides a complex representation of $x(t)$ \cite{gabor_theory_1946}. Its imaginary part $\tilde{x}(t)$ is calculated by imposing a phase shift of $\pi/2$ to the real-valued time series, which is achieved by applying a discrete Hilbert-transform. A computationally cheap way to calculate this transform is to apply a Fourier transform, multiply the positive frequency components by two, neglect the negative frequency components, and to transform back into the time-domain \cite{marple_computing_1999}.

\subsection{Stationary surrogate data serve as control}
\label{sec:surrogatedata}

%Outline
%\begin{itemize}
%\item we use AAFT surrogate time series (\Cref{fig:surrft}a) that can be interpreted as realizations of a stationary linear Gaussian stochastic process
%\item we use them as a reference when interpreting the instantaneous phase and judging stationarity
%\item surrogate time series are randomly drawn but share certain characteristics with original time series
%\item we use surrogate time series that share the Fourier transform modulus with the original time series (\Cref{fig:surrft}b), but Fourier transform argument is randomized (\Cref{fig:surrft}c)
%\end{itemize}

We use simulated surrogate time series as a reference for interpreting the linearity of the instantaneous phase. These surrogate time series are constituted by random variables that share certain characteristics with the test time series and are stationary by construction. Therefore, comparison of surrogate and test time series can reveal non-stationarity of the latter.

\begin{figure}
\includegraphics[width = 8.6cm]{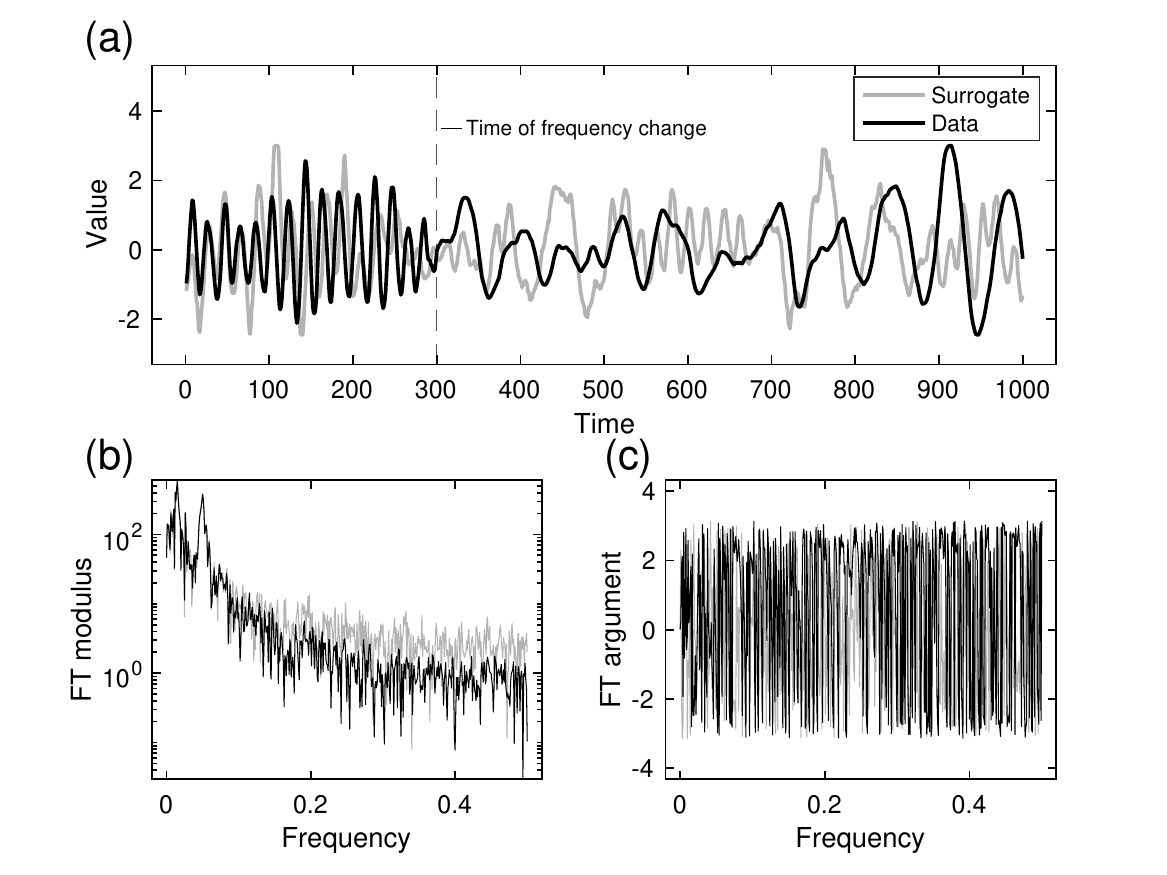}
\caption[Surrogate time series]{(a) Example time series (black) from \Cref{fig:dataphase}a  and a surrogate time series (gray). 
(b) Modulus of the Fourier-transform of the two time series in (a). (c) Argument of the Fourier-transform of the two time series in (a).}
\label{fig:surrft}
\end{figure}

The method of surrogate data \cite{theiler_testing_1992} is a procedure to test data for a feature by comparing it to simulated data that share most characteristics with the test data but do not include the feature tested for. In our case, we simulate surrogate time series (\Cref{fig:surrft}a) from which we construct the time-dependent distribution of instantaneous phases of a stationary stochastic process that approximately share the Fourier transform modulus (\Cref{fig:surrft}b) and exactly share the amplitude distribution with the test time series but have randomized Fourier transform arguments (\Cref{fig:surrft}c).

To that end, we use the amplitude adjustment Fourier transformation (AAFT) algorithm  \cite{theiler_testing_1992} to generate surrogate time series from a test time series. This algorithm works by re-ordering random numbers drawn from a normal distribution in such a way that their ranks match those of the test time series. Next, the arguments of this new time series' Fourier transform are randomized. We then back-transform the new time series into the time domain and use it as a reference for re-ordering the test time series to obtain the surrogate time series. Thereby, the null hypothesis is that of a Gaussian stationary linear process observed via an instantaneous measurement function that is invariant under translations of time \cite{theiler_testing_1992}.

In this process, the Fourier transform modulus is not exactly preserved for time series of finite length. Instead, we can observe a shift in the surrogate's Fourier transform modulus towards white noise (\Cref{fig:surrft}b). This is because an estimate of the instantaneous measurement function will not be exact for finite length of the time series, and consequently, small deviations arise that add uncorrelated random numbers to the data \cite{schreiber_surrogate_2000}. As a consequence, also the variance of the Fourier transform modulus frequency increases. %The correlation structure in the Fourier transform argument  (\Cref{fig:surrft}c) is created by the sequence of alternating high and low frequency regimes in the time series.
 
The usual problem of the method of surrogate data is that of a composite null hypothesis \cite{timmer_what_2000}. However, in our case, we account for non-Gaussianity by the usage of normally distributed random numbers in the process of generating the AAFT surrogate time series. Non-linearity can be detected e.g.\ from the correlation structure of the bi-spectrum \cite{greb_interpretation_1988} or the presence of higher harmonics in the power spectrum. There were no signs for non-linearity in the data we analyzed (\Cref{sec:application}) Therefore, the only remaining property in the null hypothesis is non-stationarity, which is also the property we want to test for.

\subsection{Confidence bands are derived from surrogate phases}
\label{sec:confbands}

%Outline
%\begin{itemize}
%\item we test a time series for stationarity by comparing its instantaneous phase to that of $\nsurr$ surrogate time series (\Cref{fig:surrphase})
%\item this is done by deriving confidence bands from the stationary surrogate time series
%\end{itemize}

We test a time series for stationarity by comparing its instantaneous phase to that of a number of surrogate time series. This is done by deriving confidence bands from the latter.

% Construction of confidence bands
%\begin{itemize}
%\item a confidence band (\Cref{fig:surrphase}) is constructed from instantaneous phase of an ensemble of $\nsurr$ surrogate time series as follows
%\item we first demean the ensemble
%\item then we calculate the $\beta$ and $1-\beta$ quantiles at each time point, they serve as preliminary upper and lower boundaries of the confidence band 
%\item we then adjust the coverage of the confidence band by iteratively increasing its size
%\item this is done by scaling upper and lower boundaries with a common factor $\gamma$ until the they contain  $\geq 100\times( 1-\alpha)\%$ of all surrogate phase curves 
%\item individual factors for upper and lower bound did not yield higher power of the test
%\item the null hypothesis of a stationary time series is rejected if the instantaneous phase of the original time series crosses a boundary 
%\end{itemize}

\begin{figure}
\includegraphics[width = 8.6cm]{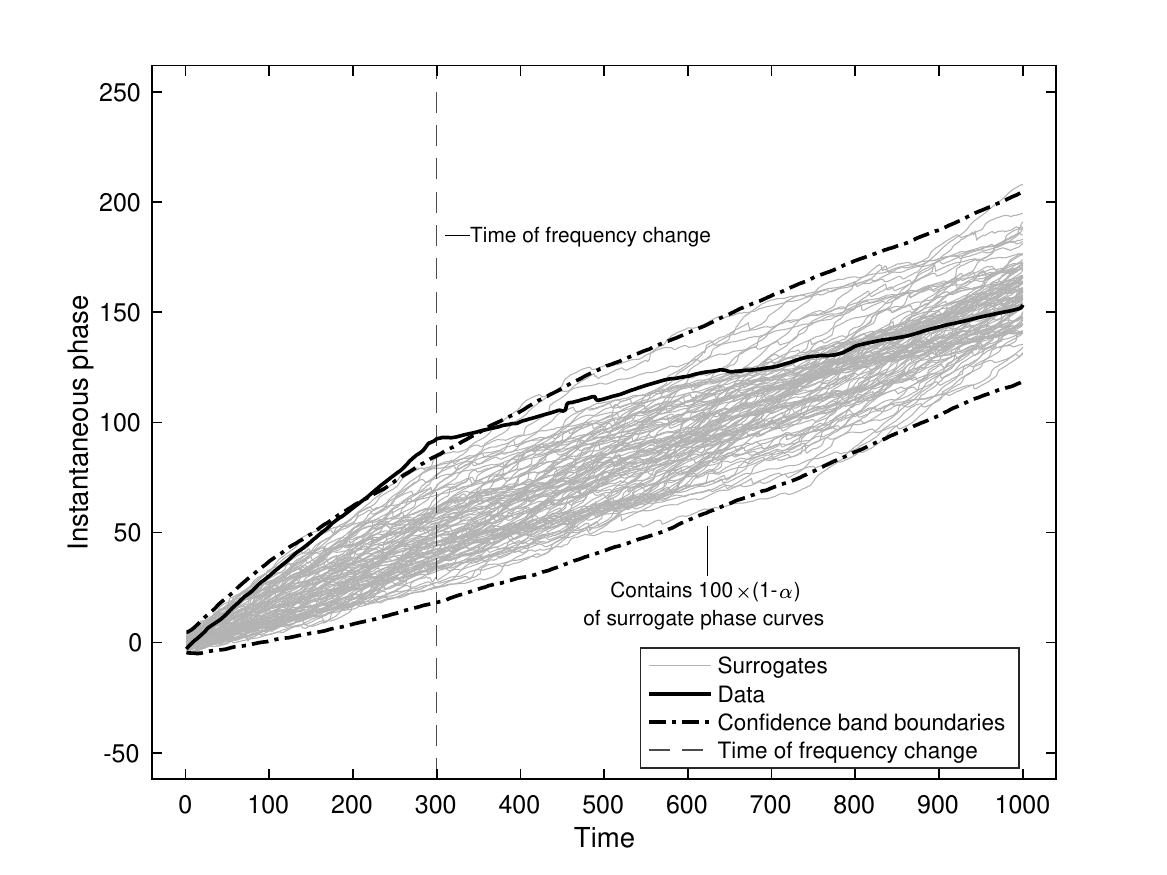}
\caption[Instantaneous phase]{Instantaneous phase of non-stationary example time series (black) from \Cref{fig:dataphase}a and of $\nsurr$ stationary surrogate time series (gray). Confidence bands for stationarity (dashed-dotted line) corresponding to a confidence level of $\alpha = 0.05$ are derived from the latter. Only a subset of surrogate phases is shown for better visibility.%Red highlighting denotes the times where the instantaneous phase of the example time series is outside the confidence band
}
\label{fig:surrphase}
\end{figure}

A confidence band (\Cref{fig:surrphase}) is constructed from an ensemble of instantaneous phases of surrogate time series as follows: We remove the mean from the ensemble and then calculate the $\beta/2 = 0.2$ and $1-\beta/2$ quantiles at each time point. These two curves serve as preliminary lower and upper boundaries, which correspond to a point-wise confidence band. They neglect the fact that a trajectory that is extremal with respect to the ensemble only for a short period of time must be excluded from the confidence band as a whole. Therefore, these boundaries are too narrow and result in a size of the test larger than the significance level $\alpha$. Size is the probability to incorrectly reject the null hypothesis if it is true, i.e.\ the probability of a Type I error. We therefore adjust the confidence band by iteratively widening its boundaries: We scale lower and upper boundaries with a common factor $\gamma$ until the they contain  $\geq 100\times( 1-\alpha)\%$ of all surrogate phase curves, i.e.\ have the correct size. The null hypothesis of stationarity is rejected if the instantaneous phase of the test time series is outside of the confidence band confined by the lower and upper boundaries at some point in time. We also tested having individual scaling factors for lower and upper boundaries. However, this did not yield a higher power of the test.

The width of the confidence band increases with time. The ensemble of instantaneous phases has a maximum variability of $2\pi$ at $t=0$. With every unwrapping at a phase jump of $\geq 2\pi$, differences in instantaneous phase between individual surrogate time series are accumulating and lead to a widening of the ensemble.

\subsection{Evaluation of size and power}
\label{sec:powerstudy}

\begin{figure}
\includegraphics[width = 8.6cm]{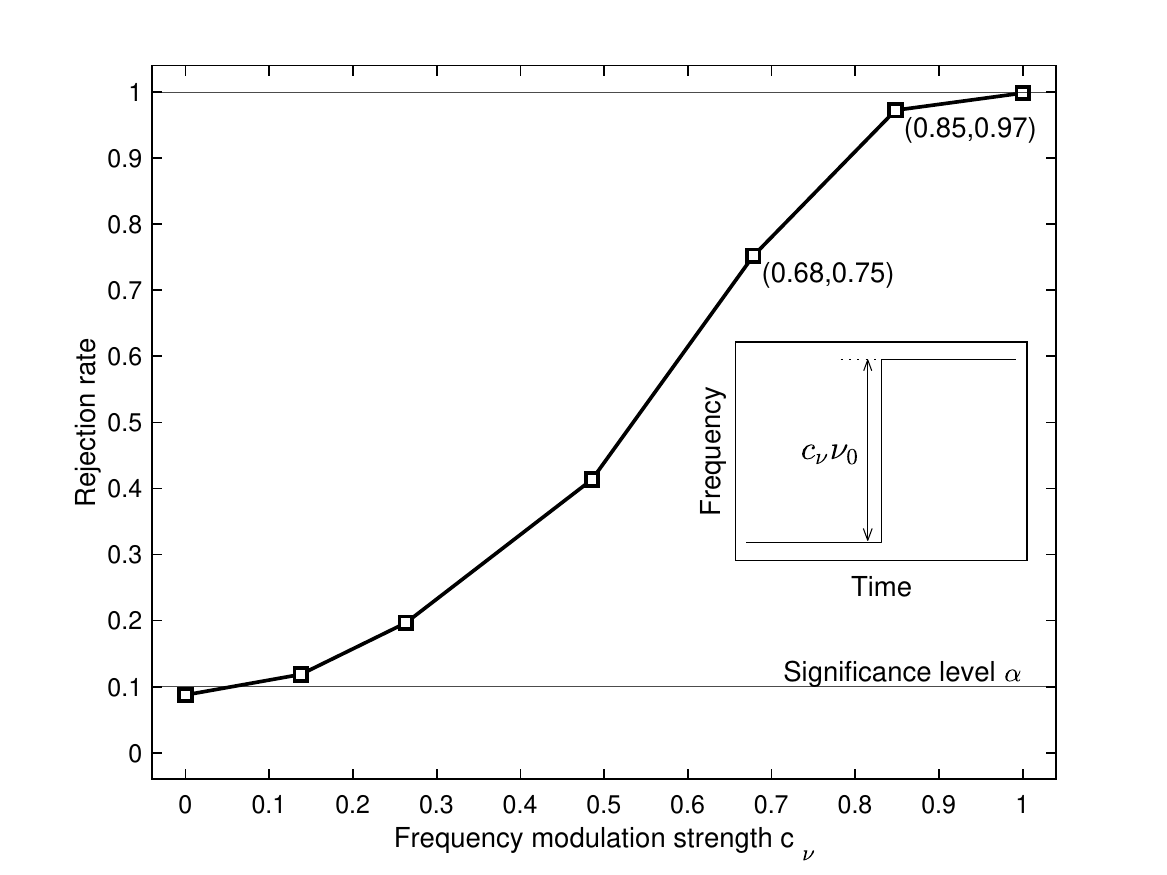}
\caption[Size and power of the test]{Evaluation of the size and power of the test. For different frequency modulation strengths $c_\nu$ (horizontal axis; \Cref{eq:stepfct}), we simulated $\nreal = 2000$ realizations of a second-order autoregressive process and counted the rate of rejections of the null hypothesis (vertical axis) of the proposed test at a significance level of $\alpha = 0.1$. The test we propose is able to identify nearly $100\%$ of non-stationary time series for $c_\nu \geq 1$ and does not lead to a fraction of false positive results larger than $\alpha = 0.1$ for $c_\nu = 0$.}
\label{fig:pott}
\end{figure}

To analyze the performance of the test we propose, we calculated size and power of the test in a simulation study. Power is the ability of the test to correctly reject the null hypothesis if it is false, i.e.\ $1-[ \text{probability of Type II error} ]$. To estimate these two quantities, we simulated $\nreal = 2000$ time series with different degrees of frequency modulation and therefore non-stationarity and counted the relative rate of rejections of the null hypothesis as a function of the frequency change (frequency modulation strength $c_\nu$; \Cref{fig:pott}).

As test data, we used realizations of an oscillating second-order autoregressive process (\Cref{eq:ar2process}) with a mixing time of $\tau = 20$ and $1100$ time points, but discarded the first $100$ time points, which corresponds to $5\tau$. This was done to make sure that correlation with the initial values decayed to a sufficiently small value. To introduce non-stationarity by modulating the frequency, we replaced $\nu$ in \cref{eq:a1} with 
% We simulated time series with a mixing time of $\tau = 20$ and $1100$ time points, but discarded the first $100$ time points, which corresponds to $5\tau$. This was done to make sure that correlation with the initial values decayed to a sufficiently small value. To introduce non-stationarity by modulating the frequency, we replaced $\nu$ in \cref{eq:a2} with 
\begin{align} 
\nu_0\gamma(t) , \quad \gamma(t) = 
\begin{cases} 
      1 & t\leq 500 \\
      2^{c_\nu} & t > 500
\end{cases}
.
\label{eq:stepfct}
\end{align}

%Results
%\begin{itemize}
%\item we chose a significance level of $\alpha = 0.1$ 
%\item beta???
%\item base frequency!!!!
%\item for stationary time series ($c_\nu = 0$), the rejection rate is almost $0.1$, corresponding to the  the significance level $\alpha=0.1$
%\item the rejection rate increases with the non-stationarity-factor $c_\nu$ and passes $80\%$ at $c_\nu=0.5$ and reaches nearly $100\%$ at $c_\nu = 1$
%\end{itemize}

With a significance level of $\alpha = 0.1$, a base frequency of $\nu_0 = 1/20$, and $\nsurr$ surrogate time series, we found that for stationary time series ($c_\nu = 0$, no frequency modulation), the rejection rate is almost $0.1$, matching the significance level. The rejection rate increases with the frequency modulation strength $c_\nu$, passes $75\%$ at a frequency increase of $2^{0.68}-1 \approx 60\% $ and reaches nearly $100\%$ at $2^1 -1= 100\%$ change relative to the base frequency (\Cref{fig:pott}).

\begin{figure*}
\includegraphics[width = 17.2cm]{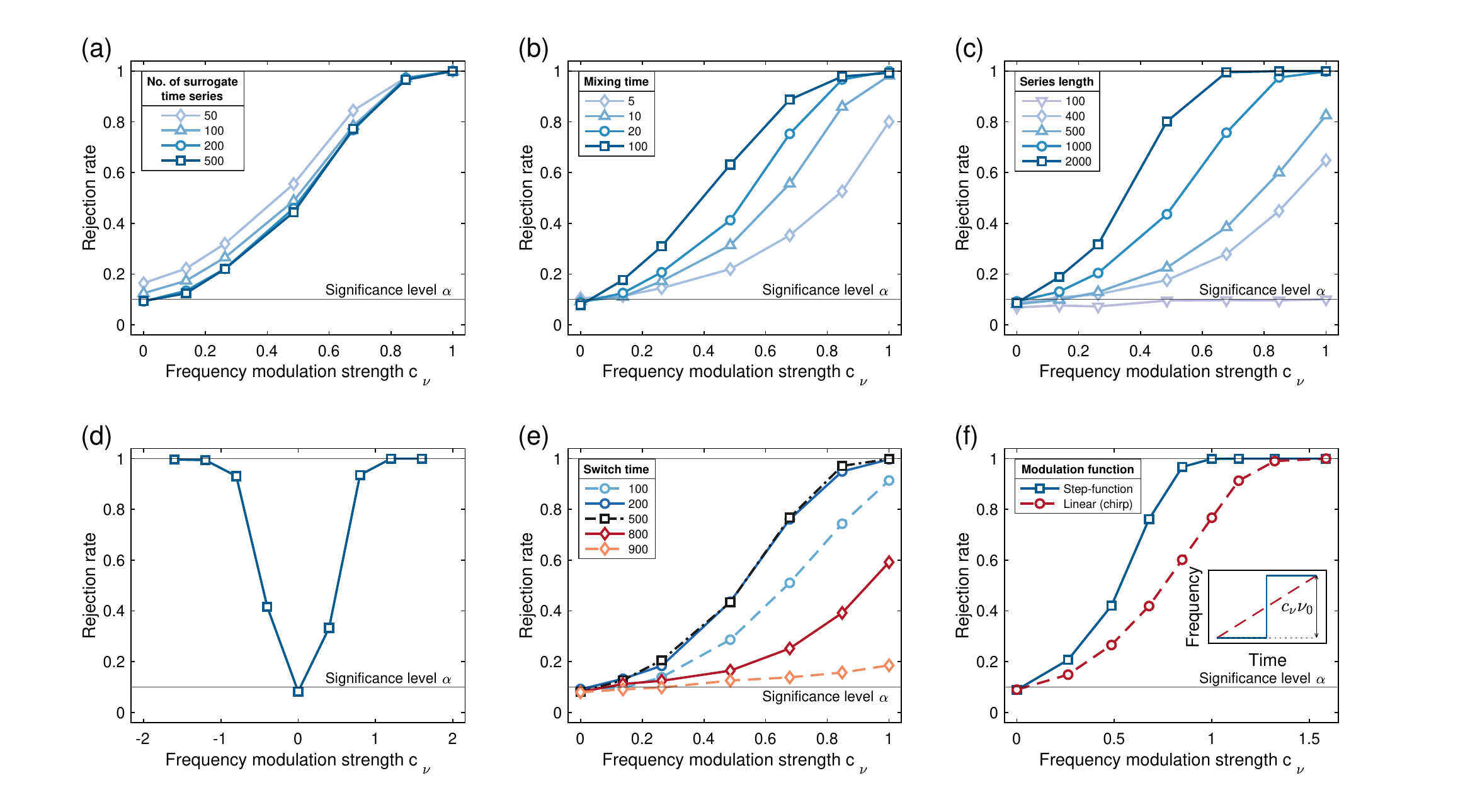}
\caption[Power in additional conditions]{We analyzed various characteristics of the simulated data and the test itself with respect to their influence on the power of the test (c.f.\ \Cref{fig:pott}) with $\nu_0 = 1/20$ and $\tau$ = 20. (a) Power curves for different numbers of surrogate time series. (b) The mixing time $\tau$ in the simulated time series was tuned. (c) The influence of time series length on the power of the test. (d) The effect of decreasing the frequency on the power of the test. (e) Power curves for different times of frequency modulation. (f) Compared with the step-function frequency modulation used in the other simulations, a chirp signal (linear modulation) with the same initial and final frequencies leads to an overall lower power. All time series used for the analyses had a length of 1000 time points if not indicated otherwise.}
\label{fig:pott_additional}
\end{figure*}

We also tested size and power of the test when varying the number of surrogate time series, mixing time $\tau$ or length of the time series. Furthermore, we evaluated the impact of changing the frequency to smaller instead of larger values or using a chirp signal instead of the step-function modulation.

%Number of surrogate time series
%\begin{itemize}
%\item we analyzed how the number of surrogate time series used to calculate the confidence band affects the power of the test
%\item using more than 100 surrogate time series does not lead to more power (\Cref{fig:pott_additional}a)
%\end{itemize}

When varying the number of surrogate time series used to calculate the boundaries of the confidence band, we found that using more than 200 surrogate time series does not substantially increase the power of the test (\Cref{fig:pott_additional}a). 

%%Mixing time
%\begin{itemize}
%\item power decreases as the mixing time goes down (\Cref{fig:pott_additional}b) because oscillations turn into noise
%\end{itemize}

%We also found that power of the test decreases when we decrease the mixing time (\Cref{fig:pott_additional}b). This is because stochastic oscillations turn into uncorrelated noise close to the extreme case $\tau = 0$. 

We also found that power of the test decreases when we decrease the mixing time (\Cref{fig:pott_additional}b). Low mixing times correspond to a fast decay of the autocorrelation function of the underlying process. In the limit case of $\tau = 0$, which represents white noise, all correlation structure including any possible oscillations and frequency modulation vanishes. The drop in power can therefore be attributed to the data itself, instead of being a shortcoming of the test we propose, and is inherent to all statistical methodology applied to the data.

%Time series length
%\begin{itemize}
%\item the longer the time series the higher the power (\Cref{fig:pott_additional}c), good power for $\geq 1000$ time points
%\end{itemize}

Concerning the length of the time series, as expected, the power of the test in general benefits from longer time series (\Cref{fig:pott_additional}c). At the same time, very short time series that span over only a few oscillations make the power decrease close to the level of significance regardless of the frequency modulation.

%Negative frequency changes
%\begin{itemize}
%\item the test we propose performs equally well if we decrease the frequency instead of increasing it as can be seen from the logarithmic representation in \Cref{fig:pott_additional}d
%\end{itemize}

The test we propose also works well when the frequency is decreasing instead of increasing. As one would expect, we find a symmetric rejection rate with respect to positive versus negative frequency changes (\Cref{fig:pott_additional}d). 

%Switch time
%\begin{itemize}
%\item power is maximized when both frequency regimes have equal length
%\item frequency switches in the first half of the time series lead to an overall higher power of the test compared to switches in the second half 
%\end{itemize}

%We observe that power of the test is maximized when the two frequency regimes have equal length (\Cref{fig:pott_additional}e). However, switching the frequency in the first half of the time series leads to an overall higher power of the test compared to switches in the second half. This is due to the widening of the confidence band for larger times (\Cref{sec:confbands}). 

We observe that power of the test is maximized when the two frequency regimes have equal length (\Cref{fig:pott_additional}e). Making the length of the two regimes unequal reduces the power of the test, because it makes the frequency modulation less pronounced, and ultimately removes the frequency modulation from the data when one of the regimes has length zero. However, switching the frequency in the first half of the time series leads to an overall higher power of the test compared to switches in the second half. This is due to fluctuations accumulating in the instantaneous phase within the process of unwrapping, which increase the width of the confidence band, as discussed in \cref{sec:confbands}.

%Linearly increasing frequency
%%\begin{itemize}
%\item we also checked how much power the test has when the frequency is increased with a linear function, i.e.\ replacing $\gamma(t)$ in \cref{eq:fm} with $\gamma(t) = c_\nu t/[\text{number of time points}]$
%\item power increases less fast with non-stationarity factor (\Cref{fig:pott_additional}e) compared to step-function frequency (\Cref{eq:stepfct}) and plateaus at around $90\%$
%\item because for high frequencies and low mixing times, oscillations become less and less clear and instantaneous phase looses track
%\item this is because it gets harder for instantaneous phase to keep track of increasingly rapid oscillations and low mixing times 
%\item when we increase the mixing time to $\tau = 60$, power reaches $100\%$ again for $c_\nu \geq 2$ 
%\end{itemize}

Finally we checked the power of the test when we use a chirp signal, i.e.\ replacing $\gamma(t)$ in \cref{eq:stepfct} with
\begin{align}
    \gamma(t) = 1+ \frac{2^{c_\nu}t}{[\text{number of time points}]} .
\end{align}
We find that the power of the test is similar to when using a step-function frequency modulation (\Cref{eq:stepfct}). However, in the latter case, the variance of frequencies is larger compared to the linear frequency increase with the same initial and final values in the chirp signal, which is why the power of the test increases less fast with $c_\nu$. 

\subsection{Application to experimental data}
\label{sec:application}
\begin{figure}
\includegraphics[width = 8.6cm]{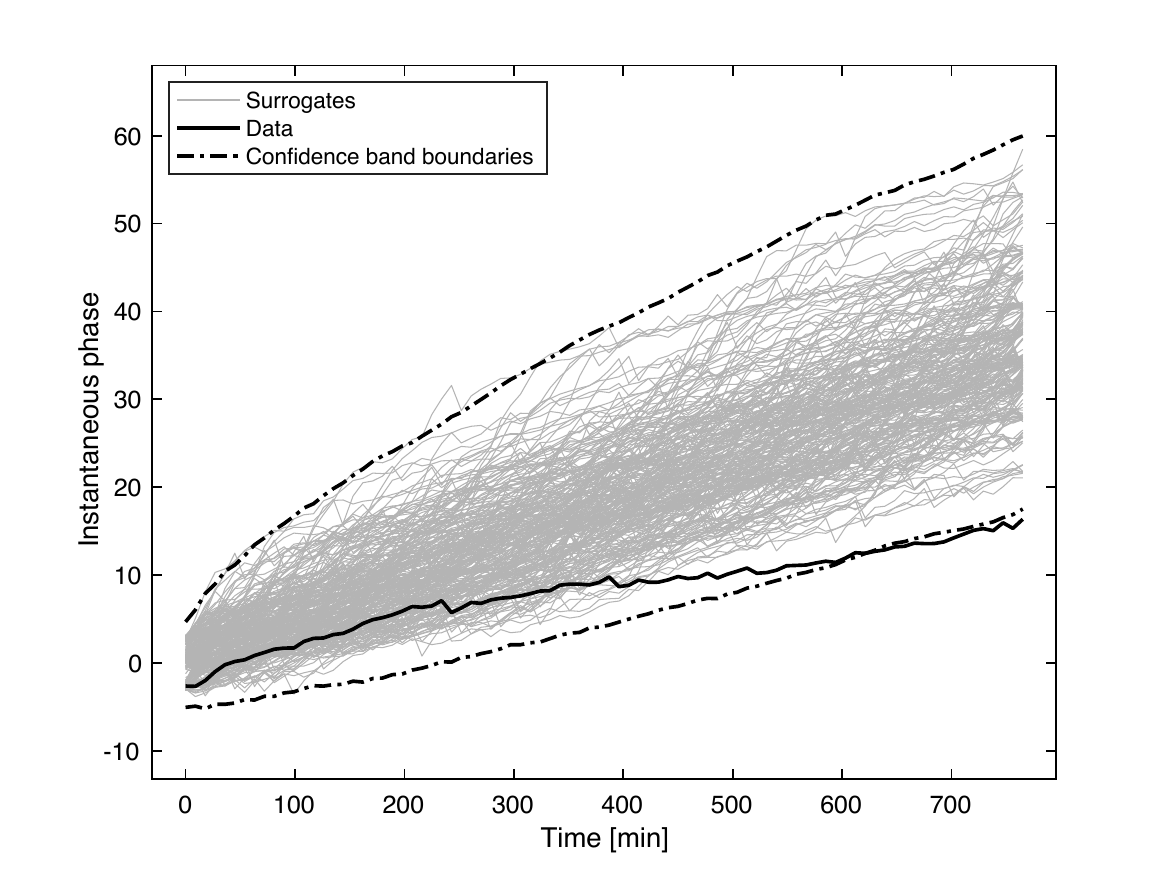}
\caption[Experimental data]{We applied our method to the experimental data from \Cref{fig:simstudy_arpars}d. Confidence bands represent a confidence level of $\alpha = 0.05$. The test we propose rejects the null hypothesis of stationarity because the instantaneous phase of the experimental data exceeds the confidence band. Only a subset of surrogate phases is shown for better visibility.}
\label{fig:siglochdata}
\end{figure}

We applied our method to time series of HES/Her family protein expression in neural stem cells of zebrafish larvae that were captured using light sheet microscopy (\Cref{fig:siglochdata}). This data features rather short time series with only 86 time points and only a few oscillations due to the nature of the biological oscillation and imaging limitations \textit{in vivo}. %In \cite{sigloch_2021}, we used a modified version of the method presented here to remove non-stationary time series from further statistical analyses. 
There were no signs for non-linearity in the power spectrum (\Cref{sec:surrogatedata}). %We find that also the improved method presented here rejects the null-hypothesis of stationarity (\Cref{fig:siglochdata}).

\section{Conclusions} \label{sec:conclusions}
We propose a new statistical test for non-stationary stochastic processes causing frequency modulation especially in short time series. We evaluated its power and size and found that modulating the frequency by $80\%$ and more relative to the base frequency leads to a correct identification of more than $97\%$ of all non-stationary time series.

%>> "Limitations, Details, How to interpret "
%"Limitations in filling gap
We tested our method with two functional forms of frequency modulation, a step-function and a linear function. In a real-world scenario however, frequency modulation might come in a different form. We hypothesize that, for example, a fast oscillatory frequency modulation might not be detected by the test we propose, because deviations in the instantaneous phase from its linear appearance average out and the instantaneous phase does not exit the confidence band. 

We demonstrated that the proposed test has power to detect frequency modulation in data that have only one frequency at any given time. The superposition of signals with different frequencies but very similar amplitudes will decrease the performance of the method, but not lead to false positive results. Also, in data with extremely low mixing times or largely instationary sections, the power of our method is subpar. However, applying a statistical method to test for frequency modulation only makes sense if there are substantial portions of the data that portray  differing frequencies, which excludes the extremal cases discussed in the simulation studies depicted in \Cref{fig:pott_additional}e and f. Concerning unknown mixing and switch times, similar simulation studies can be conducted to determine the type II error, e.g. by estimating mixing times based on the autocorrelation functions.

%---
%"limits in generalization"
The test we propose is tailored to detecting non-stationary processes in short time series that exhibit frequency modulation. Other types of non-stationarity might remain hidden to our method. Furthermore, there are no pure stationary processes in nature. Therefore, the null hypothesis will always be rejected for enough data \cite{timmer_what_2000}. However, our aim is detecting biologically relevant non-stationarity processes for which data is typically limited both in quantity and quality. %?????

%>>> title, framing: detect frequency modulation, not non-stationarity???
%---
%>> "Strength, Usefulness, The difference made"
In summary, we propose a computationally cheap and easily interpretable test for frequency modulated time series, which facilitates widespread usage and communication of results. It takes less than one second to analyze a series with 1000 time points on a state-of-the-art CPU (Intel Core i9-9880H). Additionally, there are next to no hyperparameters that have to be chosen dependent on the data except for the number of surrogate time series which shows a saturation behaviour for its influence on the power of the test. The usage of surrogate data renders the test non-parametric, removing the need for a distribution assumption that can be violated, and allows the confidence bands being calculated in a way such that the test has the correct size by construction. 

We believe that the method we propose provides a vivid meaning to non-stationarity in terms of frequency modulation and hope that this will further facilitate its application on real-world data.

%---
%"Contributions beyond"
%"Science is better now"

\section*{Code availability}
An implementation of the described method for MATLAB together with a demonstration script is available on \href{https://github.com/adrianhauber/FMtest}{https://github.com/adrianhauber/FMtest}.

\section*{Acknowledgements} \label{sec:acknowledgements}
This study was supported by the German Research Foundation (DFG) under Germany's Excellence Strategy (CIBSS - EXC-2189 - Project ID 390939984) and SFB1381. The authors acknowledge support by the state of Baden-Württemberg through bwHPC and the German Research Foundation (DFG) through grant INST 35/1134-1 FUGG.

\bibliography{bibliography.bib}

%\appendix*
%\input{sections/appendix1.tex}

\end{document}